\documentstyle[prb,aps,psfig,epsf,epsfig,twocolumn]{revtex}
\newcommand {\be}{\begin{equation}}
\newcommand {\ee}{\end{equation}}

\tighten

\begin{document}

\title{\bf A cluster algorithm for Potts models with fixed spin densities}
\author{R.P. Bikker and G.T. Barkema}
\address{Theoretical Physics, Utrecht University,
Princetonplein 5, 3584 CC Utrecht, the Netherlands}
\date{\today}

\maketitle

\begin{abstract}
A cluster algorithm is presented for the simulation of the $q$-state
Potts models in which the number of spins is conserved in each state.
The algorithm constructs Fortuin-Kasteleyn cluster configurations from
spin configurations, in a way identical to the Swendsen-Wang
algorithm; the spin assignment to these clusters is however different,
and conserves the number of spins for each state.  Compared to
traditional non-local spin-exchange algorithms, the cluster algorithm
presented here suffers less from critical slowing down, and
consequently is more efficient near the critical
temperature. 
\end{abstract}

\section{Introduction}

Before 1987, the Potts model was almost exclusively simulated by means
of the Metropolis algorithm \cite{metrop}, in which single-spin updates
are proposed and either accepted or rejected depending on the change in
energy. This algorithm works quite satisfactorily, except close to the
critical point. At the critical temperature, the correlation times
increase with system size as $L^z$, with a critical dynamic exponent
equal to or slightly above two: for the two-states Potts model (Ising
model), the critical dynamical exponent is reported to be $z=2.167 \pm 
0.001$ in two, and $z=2.02 \pm 0.02$ in three dimensions
\cite{nightingale}.  The introduction of cluster algorithms has
greatly advanced the accuracy with which critical properties of the
Potts model and many other models in statistical physics can be
studied.  The first widely used cluster algorithm was introduced by
Swendsen and Wang \cite{sw,newman}; we will describe their algorithm in
section \ref{algorithm}.  The dynamic exponent of the Swendsen-Wang
algorithm in the two- and three-dimensional Ising model is reported to
be $z=0.25 \pm 0.01$ and $z=0.54 \pm 0.02$, respectively:\cite{codd}
cluster algorithms are able to significantly reduce critical slowing
down.

To study multi-component lattice gases in the co-existence regime, for
instance to study interfaces or equilibrium crystal shapes, one has to
fix the number of particles in the lattice gas for each component,
{\it i.e.}, the spin density for each state.  One typically resorts to
spin-exchange dynamics, with the unfortunate consequence of a critical
slowing down at least as severe as experienced with the Metropolis
algorithm applied to the regular Potts model.  The usual cluster
algorithms do not conserve the spin densities. For the
conserved-order-parameter Ising model, Heringa and
Bl\"ote~\cite{heringa1,heringa2} recently introduced a cluster
algorithm, which is in spirit related to the Wolff
algorithm~\cite{wolff,newman}. It is reported to have hardly any
critical slowing down, with a dynamical exponent of $z=0.21$.  This
algorithm has not been generalized to Potts models with more than two
states.

In this paper, we present a modification of the Swendsen-Wang
algorithm, to conserve the spin densities. In the coming section, we
describe their algorithm, and introduce our modified density-conserving
cluster algorithm.  In the next section, we present measurements of the
critical dynamic exponent for our algorithm, and show its efficiency. 
The paper is concluded with a summary and conclusions, and a
discussion of future work. 

\section{Cluster algorithms}
\subsection{The Swendsen-Wang algorithm}
\label{algorithm}

The Swendsen-Wang algorithm is designed to simulate the Potts model, defined by
the Hamiltonian
\begin{equation}
H=-J \sum_{\langle i,j \rangle} \delta \left( \sigma_i, \sigma_j\right),
\label{eq:hamiltonian}
\end{equation}
in which $J$ is the coupling constant, $\delta$ denotes the Kronecker
delta function, and the summation runs over all pairs of
nearest-neighbor sites, each having a spin with value ($\sigma=1 \dots
Q$).  We use the usual symbols $N$ for the number of lattice sites, $L$
for the lateral dimension of the lattice with periodic boundary
conditions, and $\rho_i=(1/N) \sum_k \delta \left(\sigma_i,k\right)$
for the density of spins with value $i$. 

In this algorithm, the entire lattice is divided into clusters of aligned
spins, to each of which a random new value is assigned. In detail,
one step of the algorithm proceeds as follows:
\begin{enumerate}
\item Visit all nearest-neighbor pairs of lattice sites; do nothing if
the two spins are not aligned, but if they are, activate the bond
between those two sites with a probability $P_c=1-\exp(-\beta J)$,
where $\beta$ is the inverse temperature.
\item Group lattice sites that are connected by such activated bonds
into clusters.
\item Select a random new spin value for each cluster, and assign this
spin value to each of the sites constituting the cluster.
\end{enumerate}
Steps 1, 2 and 3 are to be repeated many times, to obtain a set of sample
configurations.

The proof of correctness for our density-conserving cluster algorithm
is based on that for the Swendsen-Wang algorithm, which is presented
in the remainder of this section. First we show detailed balance, next
we discuss ergodicity.

Suppose we denote the spin configuration before and after the move by
$C_a$ and $C_b$, respectively, with total energies $E_a$ and $E_b$, and
the intermediate cluster configuration $C_m$ (also known as
Fortuin-Kasteleyn representation \cite{Forkas}); furthermore, we write
the probability to move from a configuration $X$ to configuration $Y$
as $T(X\rightarrow Y)$.  Then, the probability to move from a spin
configuration $C_a$ to a cluster configuration $C_m$ is a product with
factors $P_c$ over all nearest-neighbor pairs of spins that are
connected, times a product with factors $1-P_c$ over all aligned
nearest-neighbor pairs of spins that are disconnected:
\begin{equation}
T(C_a\rightarrow C_m)=\prod_{
  \begin{array}{ccc}
    \langle i,j\rangle\\
    \sigma_i^{(a)}=\sigma_j^{(a)}\\
    \mbox{$i$, $j$ conn.}
  \end{array}}
  \hskip -0.4cm \Bigl( P_c \Bigr) \hskip -0.2cm 
  \prod_{
  \begin{array}{ccc}
    \langle i,j\rangle\\
    \sigma_i^{(a)}=\sigma_j^{(a)}\\
    \mbox{$i$, $j$ disconn.}
  \end{array}}
  \hskip -0.5cm \Bigl( 1-P_c \Bigr)
\end{equation}
and a similar expression for $T(C_b\rightarrow C_m)$. Since spins that
are connected, are necessarily aligned both before and after the move,
the first product on the right hand side is equal in $T(C_a\rightarrow
C_m)$ and $T(C_b\rightarrow C_m)$. All factors in the second product
on the right hand side dealing with pairs of spins that are aligned
both before and after the move are also equal in $T(C_a\rightarrow
C_m)$ and $T(C_b\rightarrow C_m)$. That leaves in the ratio of the
transition rates only the factors dealing with disconnected pairs of
spins that are aligned either in configuration $C_a$, or in
configuration $C_b$, but not both. The ratio of the transition rates
$T(C_a\rightarrow C_m)$ and $T(C_b\rightarrow C_m)$ therefore reduces
to  
\begin{equation}
\frac{T(C_a\rightarrow C_m)}{T(C_b\rightarrow C_m)}=
\prod_{
  \begin{array}{ccc}
    \langle i,j\rangle\\
    \sigma_i^{(a)}=\sigma_j^{(a)}\\
    \sigma_i^{(b)}\neq \sigma_j^{(b)}
  \end{array}}
  \hskip -0.4cm \Bigl( 1-P_c \Bigr) \hskip 0.15cm \Big/ \hskip -0.6cm 
\prod_{
  \begin{array}{ccc}
    \langle i,j\rangle\\
    \sigma_i^{(a)}\neq \sigma_j^{(a)}\\
    \sigma_i^{(b)}=\sigma_j^{(b)}
  \end{array}}
  \hskip -0.5cm \Bigl( 1-P_c \Bigr).
\end{equation}
Using that $\log(1-P_c)=-\beta J$, in combination with some rewriting, we
obtain for the logarithm of this ratio
\begin{eqnarray}
& &\log(T(C_a\rightarrow C_m))-\log(T(C_b\rightarrow C_m)) \nonumber \\
% &=& \sum_{
% \begin{array}{ccc}
% \langle i,j\rangle\\
% \sigma_i^{(a)}=\sigma_j^{(a)}\\
% \sigma_i^{(b)}\neq \sigma_j^{(b)}
% \end{array}}
% \hskip -0.5cm \Bigl( -J \Bigr) -
% \sum_{
% \begin{array}{ccc}
% \langle i,j\rangle\\
% \sigma_i^{(a)}\neq \sigma_j^{(a)}\\
% \sigma_i^{(b)}=\sigma_j^{(b)}
% \end{array}}
% \hskip -0.5cm \Bigl( -J \Bigr) \nonumber \\
&=& -\beta J \sum_{\langle i,j\rangle}
  \left[ \delta \left(\sigma_i^{(a)},\sigma_j^{(a)}\right)
        -\delta \left(\sigma_i^{(b)},\sigma_j^{(b)}\right) \right].
\end{eqnarray}
As can easily been seen from the Hamiltonian eq.
(\ref{eq:hamiltonian}), this is equal to $-\beta(E_a-E_b)$.  Since
$T(C_m\rightarrow C_a)=T(C_m\rightarrow C_b)=2^{-n}$ where $n$ is the
number of clusters in $C_m$, detailed balance follows:
\begin{eqnarray}
\frac{T(C_b\rightarrow C_a)}{T(C_a\rightarrow C_b)} & = &
\frac{T(C_b\rightarrow C_m)\cdot T(C_m\rightarrow C_a)} 
     {T(C_a\rightarrow C_m)\cdot T(C_m\rightarrow C_b)}\nonumber\\
  &=&\exp\left(-\beta(E_a-E_b)\right).
\label{db}
\end{eqnarray}
In addition to obeying detailed balance, the algorithm is ergodic,
since there is a finite probability that in a given move all clusters
will contain one site only, to which any value can be assigned. Since
this algorithm is ergodic and satisfies detailed balance, it is
guaranteed that eventually these sample configurations are drawn from
the Boltzmann distribution for the regular Potts model. The densities
$\rho_i$ are not conserved in the Swendsen-Wang algorithm. 

\subsection{Density-conserving cluster algorithm}

The topic of this paper is to present a modification to this algorithm,
that ensures the conservation of the densities. This modification is
made in step 3, in which the new spin values are assigned: rather than
assigning random spin values to each cluster, we redistribute spin
values over the clusters while conserving the spin densities.  As for
the original Swendsen-Wang algorithm, the general idea is a two-step
approach, $C_a \rightarrow C_m \rightarrow C_b$, where all the
energetics required for obtaining detailed balance are incorporated in
the construction of the clusters, and detailed balance is achieved by
conservation of the property $T(C_m\rightarrow C_a)=T(C_m\rightarrow
C_b)$. 
% $C_m$ here is a labeled Fortuin-Kasteleyn type configuration,
% {\em i.e.}, a cluster configuration where we remember the spin-values
% of the clusters.

The first step towards such an algorithm is to devise an elementary
move. The move we are looking for, is identifying one set of aligned
clusters with spin value $q_1$ and another such set with spin value
$q_2 \neq q_1$ with exactly the same area (number of sites), and then
exchanging the spin values $q_1$ and $q_2$.

How do we identify such sets? First of all, for each spin value
$i=1\dots Q$ we group all clusters with spin value $i$ into the set
$S_i$. Next, within each set we list these clusters in a random order,
and keep track of the cumulative area. Every time that in two sets the
same value for the cumulative area occurs, we have found an {\it
exchange point}. If the spin values are exchanged in all clusters up to
the exchange point, while the original spin values in all other
clusters are conserved, the spin values of two sets of clusters are
exchanged without violation of the spin density conservation. 

Unless extra measurements are taken, an algorithm based on these
elementary moves will not obey detailed balance: the probability of
occurrence for an exchange point is not necessarily equal before and
after the cluster exchange. We denote the total number of clusters with
spin-value $q$ before the exchange takes place as $n_q$.  Suppose that
the exchange takes place between clusters with spin $1$ and $2$,
and that the number of clusters with spin $1$, $2$ that are to be
exchanged is $a_1$ and $a_2$, respectively, while the number of
clusters with spin $1$, $2$ that are not to be exchanged is
$n_1-a_1$ and $n_2-a_2$, respectively.  The likelihood that there is an
exchange point exactly between these sets of clusters is then equal to
\begin{equation}\label{eq:TR}
T(\rightarrow)=
\left[ \left(n_1 \atop a_1 \right) \left(n_2 \atop a_2 \right) \right]^{-1}
\end{equation}
while after the exchange, this probability becomes 
\begin{eqnarray}\label{eq:TL}
T(\rightarrow) & = &
\left[ 
\left(n_1' \atop a_2 \right)
\left(n_2' \atop a_1 \right) 
\right]^{-1} \nonumber \\
& = & \left[ 
\left({a_2+n_1-a_1} \atop a_2 \right)
\left({a_1 + n_2 -a_2}\atop a_1 \right) 
\right]^{-1} 
\end{eqnarray}
To restore detailed balance, it suffices to introduce a Metropolis
acceptance ratio:
\begin{equation}
P_a=\min \left[ 1, \frac{n_1'!\cdot n_2'!}
                              {n_1!\cdot n_2!} \right].
\label{eq:flip}
\end{equation}
Once this acceptance probability is included, the elementary move
can be used for a correct algorithm, since for two configurations
$X$ and $Y$, we now restored the property $T(C_m\rightarrow
X)=T(C_m\rightarrow Y)$. 

In an actual implementation,
the total procedure is to make for each spin value a cumulative list of
clusters, where the clusters are placed in a random order. Next,
all exchange points are identified, the corresponding exchanges are
accepted with the
probability as given in eq.~(\ref{eq:flip}). It can be
verified that also for the concatenation of exchange points, the
product over all exchange points of the ratio of forward and backward
acceptance probabilities, as given in eq.~(\ref{eq:flip}), equals
\be
\prod_{q=1}^Q \left(n_q'!\right) / \prod_{q=1}^Q \left(n_q!\right),
\ee
which exactly cancels the ratio of the number of ways in which the
clusters can be sorted, i.e.\ the ratio of selection
probabilities in forward and backward direction. 
Consequently $T(C_m \rightarrow C_a) = T(C_m \rightarrow C_b)$.

The density-conserving algorithm is ergodic for the same reason that
the Swendsen-Wang algorithm is ergodic: there is a finite probability
that all clusters contain one site only, and then each of these can
obtain any spin value (under the constraint on the densities).

Having shown that the basic steps of our algorithm are correct we
will now summarize the procedure in the form of a step-wise
algorithm. Steps number 1 and 2 of the Swendsen-Wang algorithm remain
unchanged. Step 3 becomes: 
\begin{enumerate}
\item[3a.] For each state $q$, list all clusters with this spin-value
in list $S_q$, in a random order.

\item[3b.] Order the lists with respect to the total area of their
not-yet-assigned clusters. Use a random order for lists with equal
such areas. If the first two lists (those with the largest and
next-largest areas) are equal, exchange their colors with the
probability as given by eq.~(\ref{eq:flip}). Select one cluster from
the first list and assign to it a new color. Update the ordering and
repeat this step until spin values are assigned to all clusters. 
\end{enumerate}

Note that the computational effort required for step 3 scales with the
total number of clusters $n = \sum_q n_q$, whereas step 2 scales with
the number of spins $N$ in the system. Since $n \ll N$, step 3 is
repeated $N/(2n)$ times for each time step 2 is performed, and we still
have an implementation in which the computational effort per sweep
scales linearly with the number of sites; this greatly decreases the
auto-correlation time. It actually also reduces the dynamical critical
exponent $z$, since the ratio $N/n$ varies with the system size.

\begin{figure}
\begin{center}
\epsfxsize 4cm
\epsfbox{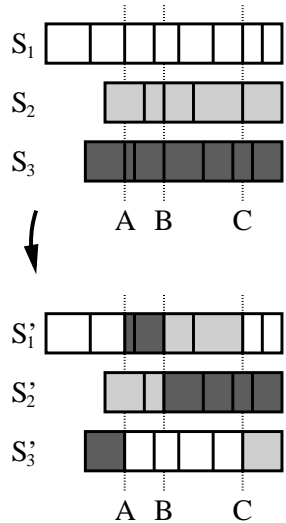}
\caption{ Assignment of new spin values to the clusters in the
three-states Potts model. The upper and lower part are the situations
before and after cluster assignment, respectively. The length of each
bar corresponds to the total mass $\rho_iN$ in the set $S_i$. The
different shades indicate different spin values before the
assignment. The thick lines separate subsequent clusters in each
list. The dotted lines indicate the exchange points $A, B$ and $C$,
where two of the masses coincide and after which the corresponding
spin values are exchanged with the probability given in
eq.~(\ref{eq:flip}).  
}
\label{sketch}
\end{center}
\end{figure}

\section{Computational properties}

\begin{figure}
\epsfxsize 8cm
\epsfbox{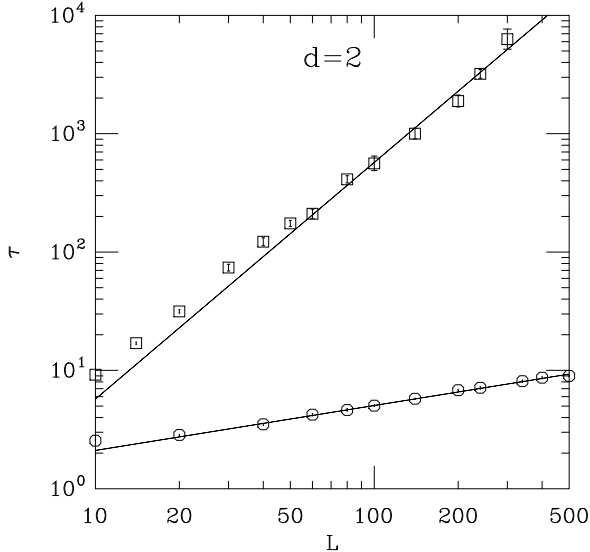}
\caption{correlation time $\tau$ as a function of linear system size
$L$ for the two-dimensional Ising model, for spin-exchange dynamics
(squares) and the magnetization-conserving cluster algorithm (circles).
The lines have exponents of $z=2$ and $z=0.38$.}
\label{twodim}
\end{figure}

\begin{figure}
\epsfxsize 8cm
\epsfbox{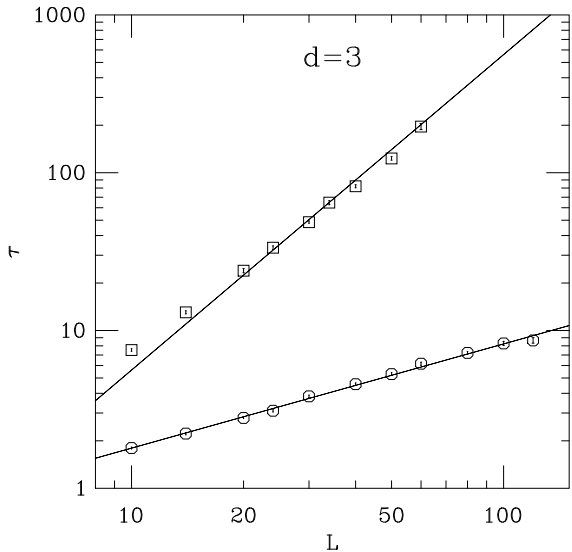}
\caption{correlation time $\tau$ as a function of linear system size
$L$ for the three-dimensional Ising model, for spin-exchange dynamics
(squares) and the magnetization-conserving cluster algorithm (circles).
The lines have exponents of $z=2$ and $z=0.66$.}
\label{threedim}
\end{figure}

In order to compare the efficiency of the density-conserving cluster
algorithm presented above with that of non-local spin-exchange
(Kawasaki\cite{Kawa}) dynamics, we have computed the energy
autocorrelation times in the Ising model at critical temperature and
equal spin densities, for several system sizes. Figures~\ref{twodim}
and~\ref{threedim} show the correlation times as a function of the
linear system size $L$ of the two- and three-dimensional Ising model,
respectively, both at their critical point. For all data points the
correlation time $\tau$ was obtained from a least-squares fit of the
form $e^{-t/\tau}$, to the energy autocorrelation function. For the
spin-exchange algorithm, these fits were done in the region where the
autocorrelation drops from $e^{-1}$ to $e^{-2}$; for the cluster
algorithm, we fitted in a broader region (from 1 to $e^{-3}$), in
order to have enough points to fit. All runs where started from a
random configuration, which was thermalized over a time varying from
6000 MCS to 60000 MCS.  In order to generate enough statistics, the
total length of the runs was set to ten times the thermalization time.
The statistical errors were determined by repeating each run 10 to 50
times. 

As expected, we find that the cluster algorithm suffers significantly
less from critical slowing down and clearly outperforms spin-exchange
dynamics at physically interesting lattice sizes in both two and three
dimensions.  Since one move in our cluster algorithm takes an amount
of CPU-time comparable to what is required for one sweep in the
non-local spin exchange, our cluster algorithm outperforms non-local
spin exchange by one or two orders of magnitude, depending on the
system size. 

For the non-local spin-exchange algorithm, we find a critical dynamic
exponent of $z=2.0$ in both two and three dimensions. This is in good
agreement with the exponents of the three-dimensional non-conserving 
Metropolis algorithm ($z=2.02 \pm 0.02$) but not with the critical
exponent for the two-dimensional non-conserving Metropolis algorithm
($z=2.167 \pm 0.001$). Perhaps this is an indication that the
conservation of the order parameter affects the critical dynamical
exponent, but our statistics are not conclusive. 

For the critical dynamic exponent for our new density-conserving cluster
algorithm, we find values of $z=0.38 \pm 0.01$ in two, and $z=0.66
\pm 0.02$ in three dimensions. These values are both slightly larger
than non-conserved Swendsen-Wang values ($z=0.25 \pm 0.01$ and $z=0.54
\pm 0.02$ for two and three dimensions respectively).

\section{Summary and future work}

We have presented a density-conserving cluster algorithm for the Potts
model.  This algorithm is only moderately sensitive to critical slowing
down:  its dynamic critical exponent is found to be $z=0.38 \pm 0.01$
and $z=0.66 \pm 0.02$ for the two- and three-dimensional two-states
Potts model, respectively. It outperforms the traditional algorithm,
non-local spin exchange, by one or two orders of magnitude.

In future research, we will use this algorithm to study wetting
properties, where the wetting takes place at a curved interface between
two co-existing phases; such non-flat interfaces arise for instance
between a droplet and a surrounding fluid. Other future applications
will include the study of line tension between three co-existing
phases, and equilibrium shapes in multi-component mixtures.

\section*{Acknowledgments}

Useful discussion with Henk van Beijeren and Matthieu Ernst is
gratefully acknowledged.

\end{document}